\documentclass[pre,twocolumn,superscriptaddress,amsmath,amssymb,floatfix]{revtex4-2}
\usepackage{graphicx}
\usepackage{xcolor}
\usepackage{xurl}
\usepackage{bm}
\usepackage[colorlinks=true,linkcolor=blue,citecolor=blue,urlcolor=blue]{hyperref}

\newcommand{\E}{\mathbb{E}}
\newcommand{\avg}[1]{\langle #1\rangle}

\begin{document}

\title{Condensation and optimal targeting of repeated forcing in heterogeneous networks}

\author{Jose Herrera-Diestra}
\email{jherreradiestra@tarleton.edu}
\affiliation{Department of Mathematics, Tarleton State University, Stephenville, Texas 76402, USA}

\date{\today}

\begin{abstract}
Many networks receive repeated external forcing through a subset of entry
nodes, for example infections imported from another population, information
seeded by a campaign, or invasive species arriving through ports. We study how
concentrating such forcing on important entry nodes, with seeding probability
$\pi_i\propto w_i^{\gamma}$, changes the size of the resulting cascades in a stochastic SIR
model on bipartite networks. The response to one seeding event is linear in the
degree of the seeded node, so the relative gain from targeting does not depend
on transmissibility. The targeting distribution is a Gibbs measure. For
importance tails $p(w)\sim w^{-a}$ and responses growing as $w^{\eta}$, the targeted
response diverges at $\gamma_1=a-1-\eta$ and the forcing condenses on a few nodes at
$\gamma_2=a-1$. Under repeated forcing, the burden-maximizing exponent $\gamma^*$ is
strictly positive, and in large populations depletion of nodes already reached
bounds it by the condensation point, $\gamma^*\le\gamma_2$. The optimum decays as $\gamma^*\simeq K/x$ for intense forcing, where $x$ is the
expected number of seeding events per entry node, and approaches the
condensation point as $\gamma_2-\gamma^*\simeq(a-1)/\ln(1/x)$ for sparse forcing. A cut-off of the
importance distribution rounds the transitions in populations smaller than a
crossover size $n_c$ and removes them in larger ones. Networks calibrated to
survey data on sexual partnerships have between $0.08\,n_c$ and $0.8\,n_c$ entry nodes,
which places them in the finite-size regime, where the asymptotic results hold
only qualitatively.
\end{abstract}

\maketitle

\section{Introduction}
Epidemics on networks are usually studied from a fixed initial condition or
from a single seed \cite{newman2002spread,kenah2007second,pastorsatorras2015epidemic}.
Many systems are instead driven repeatedly from outside. Infections are
imported from another population while an external epidemic lasts, campaigns
seed information into a social network, and invasive organisms arrive through a
limited set of ports. In such systems two quantities matter, how much forcing
occurs and where it enters. The entry points are often selected by the network
itself, because nodes with many internal contacts also receive more external
contact, as when imported cases arrive preferentially at highly connected
airports \cite{colizza2006role}.

The location of seeds affects spreading. Influence maximization asks which set
of nodes to seed once so that a cascade is as large as possible
\cite{kempe2003maximizing,morone2015influence}, and many studies rank single
spreaders by degree, $k$-shell index or percolation-based centralities
\cite{kitsak2010identification,radicchi2016leveraging,lu2016vital,erkol2019systematic}.
Seeding hubs accelerates spreading, and on geometric networks it can produce
smaller final sizes than random seeding \cite{odor2021switchover}.
Degree-biased initial conditions have exact final-size theories
\cite{miller2014epidemics}, coupled networks can sustain epidemics that neither
layer sustains alone \cite{saumell2012epidemic,dickison2012epidemics}, and
spillover between layers and its optimization have been studied numerically
\cite{das2024sir,pan2019optimizing}. Models with spontaneous infection include
external forcing but distribute it uniformly over nodes
\cite{vanmieghem2012epidemics}. To our knowledge, existing theory does not
describe repeated, time-varying forcing whose distribution over entry nodes is
tuned continuously, nor how that
distribution interacts with the response of the entry nodes, which in many
network dynamics scales as a power of the node degree
\cite{barzel2013universality,hens2019spatiotemporal}.

Concentrating the forcing with weights $w_i^\gamma$ defines a Gibbs measure,
$\pi_i\propto e^{\gamma\ln w_i}$, with inverse temperature $\gamma$ and random energies $-\ln w_i$, the
object studied by the theory of disordered systems
\cite{derrida1981random,mezard1987spin,mezard2009information}. Sums of
heavy-tailed variables condense when a single term dominates
\cite{derrida1987statistical,bouchaud1997universality,derrida1997random,majumdar2005nature,vezzani2019single},
and sums of random exponentials have critical points at which the law of large
numbers and the central limit theorem fail \cite{benarous2005limit}. These
limit laws are known, and our results concern their consequences
for a cascade driven through the measure. The response exponent of the nodes
moves one critical point and leaves the condensation point unchanged, and
depletion of nodes already reached turns condensation from a gain into a loss,
which sets a finite optimal concentration in large populations.

Here we introduce a targeting exponent $\gamma$, with seeding probability
$\pi_i\propto k_i^\gamma$ over entry nodes, and study a stochastic SIR process on bipartite
networks driven by repeated forcing. The motivating example is the importation
of a sexually transmitted pathogen into a heterosexual contact network through
bisexual men, a setting in which heavy-tailed partner numbers affect
transmission \cite{endo2022heavy}. The pathogen is hypothetical and the results
do not depend on it. We show that (i) the response to a single seeding event is
linear in the degree of the seeded node, so the relative gain from targeting
does not depend on transmissibility; (ii) the targeting distribution is a Gibbs
measure that, for degree tails $p(k)\sim k^{-a}$, has a divergence of the targeted
degree at $\gamma_1=a-2$ and a condensation transition at $\gamma_2=a-1$; (iii) under
repeated forcing, the burden-maximizing exponent $\gamma^*$ is positive and, in large
populations, depletion of already-seeded nodes bounds it by the condensation
point, with $\gamma^*\simeq K/x$ for intense forcing and
$\gamma_2-\gamma^*\simeq(a-1)/\ln(1/x)$ for sparse forcing; (iv) these results hold for any node
importance and response exponent $\eta$, with $\gamma_1=a-1-\eta$; and (v) a degree cut-off
removes the transitions above a crossover size $n_c$, which locates finite real
populations relative to the transition.

\section{Model}\label{sec:model}
\begin{figure}[t]
\includegraphics[width=\columnwidth]{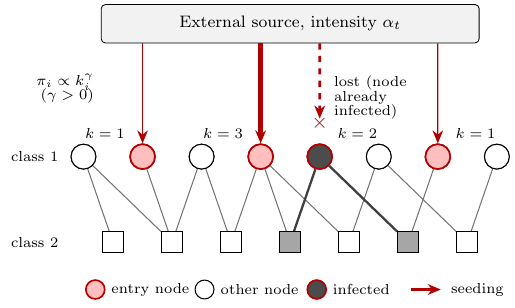}
\caption{Model. Entry nodes (red) are a subset of class 1 of a bipartite
network. An external source seeds them with probabilities $\pi_i\propto k_i^\gamma$, so for
$\gamma>0$ high-degree entry nodes receive more seeding events (arrow width). A
seeding event aimed at an entry node that is already infected or recovered
produces no infection (dashed arrow), which causes depletion. Each successful
seeding starts an SIR cluster in the network (dark nodes and edges). Schematic
drawn in TikZ with the assistance of Claude Opus 5.5 and checked by the author.}
\label{fig:schematic}
\end{figure}

\paragraph{Network.} A static bipartite configuration network has $N_1$ nodes of
class 1 and $N_2$ of class 2 \cite{newman2002spread,gomezgardenes2008spreading} with degree distributions $p_k$ and $q_k$ and mean
excess degrees $\kappa_s=\avg{k(k-1)}_s/\avg{k}_s$, $s=1,2$. Entry nodes are a subset $B$ of
class 1, $n=|B|$, chosen independently of degree among nodes with $k\ge1$, so their
degrees follow the class-1 law. For a tail $p_k\sim k^{-a}$, $\kappa_1$ is finite only for
$a>3$, which the degree model therefore requires.

\paragraph{Targeted forcing.} On day $t=1,\dots,T$ each susceptible entry node $i$ is
seeded (Fig.~\ref{fig:schematic}) with probability $1-\exp(-\beta_{\mathrm{ext}}\alpha_tn\pi_i)$, where $\alpha_t\in[0,1]$ is the
relative intensity of the external source and
\begin{equation}\label{eq:pi}
\pi_i(\gamma)=\frac{k_i^{\gamma}}{\sum_{j\in B}k_j^{\gamma}}.
\end{equation}
Negative $\gamma$ favors low-degree entry nodes, $\gamma=0$ is uniform and positive $\gamma$
favors hubs. In the epidemic application nobody chooses $\gamma$. It measures how
strongly external contact increases with the number of internal partners, as
expected when sexual activity is correlated across partnership types. The
burden-maximizing exponent $\gamma^*$ is then the worst-case allocation of
importations. For a seeding campaign it is the best allocation of seeds.
Because $\sum_i\pi_i=1$, the total daily hazard
$\beta_{\mathrm{ext}}\alpha_tn$ summed over entry nodes does not depend on $\gamma$, so $\gamma$ changes
where the forcing goes and not how much of it there is. An external contact can
infect an entry node only while that node is susceptible. If the forcing
selects an entry node that is already infected or recovered, the contact
produces no infection and is not reassigned to another entry node. As more
entry nodes are reached, a growing share of the forcing therefore produces no
new infection, an effect we call depletion. We write $c=\beta_{\mathrm{ext}}n\sum_t\alpha_t$ for the
expected number of seeding events in the absence of depletion and $x=c/n$ for
the number per entry node. The results depend on the forcing profile $\alpha_t$ only
through $c$. In simulations we use a 17-week incidence curve from the 2022 mpox
outbreak in New York City \cite{nycdohmh2022mpox}.

\paragraph{Dynamics.} In the discrete-time SIR model, each day every infectious node infects
each susceptible partner with probability $p$ and then recovers with
probability $r$. The per-edge transmissibility is
$\tau=(1-r)p/(p+r-pr)$ \cite{SM}, and $R=\tau\sqrt{\kappa_1\kappa_2}$.
We neglect overlaps between clusters started by different seeding events.
The resulting relative error is of the order of the attack fraction $C/(N_1+N_2)$,
small when entry nodes are a small fraction of the population (2\% of men
here). In simulations it reaches 10\% only under the most intense forcing tested
\cite{SM}. Results are derived in the locally tree-like approximation, with the
assumptions listed in the Supplemental Material \cite{SM}, which also gives all
proofs and the simulation design.

\section{Results}\label{sec:results}

\subsection{Linear response and targeting gain}
\begin{figure}[t]
\includegraphics[width=\columnwidth]{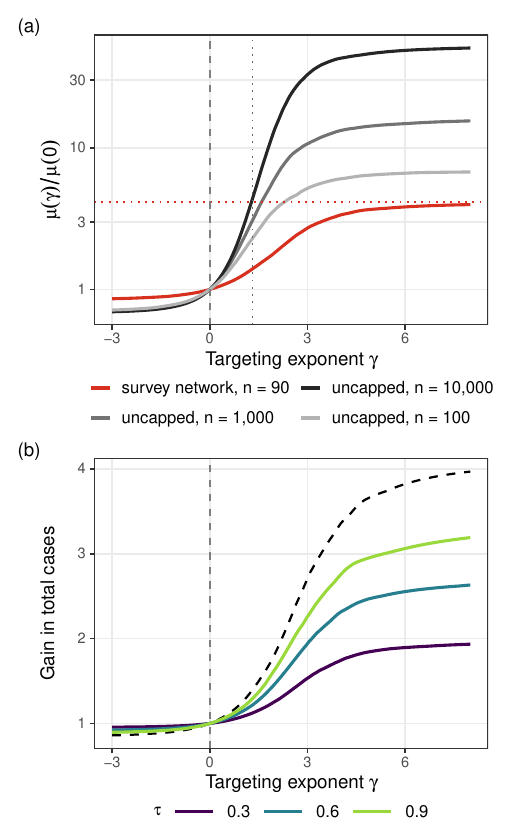}
\caption{Targeting gain per seeding event. (a) Gain in secondary cases
$\mu(\gamma)/\mu(0)$, which does not depend on transmissibility. The red curve is the
median over 100 survey-calibrated networks with about 90 entry nodes each, and
the red dotted line the median bound $k_{\max}/\mu(0)$. Grey curves are medians over
200 samples of $n=10^2$, $10^3$ and $10^4$ entry nodes drawn from the same survey law
without a degree cap. Vertical lines mark $\gamma=0$ (dashed) and $\gamma_1=a-2=1.31$
(dotted). (b) Gain in total cases on the survey-calibrated networks for
$\tau=0.3$, 0.6 and 0.9, with the gain in secondary cases (dashed) for comparison.}
\label{fig:gain}
\end{figure}

Below threshold ($R<1$), one seeding event into a node of degree $k$ produces
on average
\begin{equation}\label{eq:sk}
s(k)=Ak,\qquad A=\frac{\tau(1+\tau\kappa_2)}{1-R^2}
\end{equation}
secondary cases. With $\mu(\gamma)=\sum_i\pi_ik_i$, the expected degree of the seeded
node, the gain from targeting in secondary cases is $\mu(\gamma)/\mu(0)$,
independent of $\tau$ and of the network beyond the entry nodes, and bounded
by $k_{\max}/\mu(0)$. Near threshold the absolute effect of targeting diverges as
$(1-R^2)^{-1}$ while the relative effect stays bounded (Fig.~\ref{fig:gain}).
Equation~\eqref{eq:sk} and the probability of no secondary case agree with
simulations over network ensembles for $R\lesssim0.9$ \cite{SM}.

\subsection{Targeting as a Gibbs measure with two transitions}
Writing $\pi_i=e^{\gamma\ell_i}/Z$ with $\ell_i=\ln k_i$ makes the targeting a Gibbs
measure with inverse temperature $\gamma$ and energies $-\ell_i$
\cite{derrida1981random}. For a power-law tail, $\ell$ is exponentially distributed
with rate $a-1$, so $\pi$ is the equilibrium measure of Bouchaud's trap model,
whose glass transition at inverse temperature $a-1$ is the condensation point
$\gamma_2$ below \cite{bouchaud1992weak,bouchaud1995aging}. For entry-node degrees with tail $p(k)\sim k^{-a}$, the
weights $k^\gamma$ have tail index $(a-1)/\gamma$. Two transitions follow as $n\to\infty$.
At $\gamma_1=a-2$ the targeted degree $\mu=\overline{k^{\gamma+1}}/\overline{k^{\gamma}}$ starts to grow with $n$,
with exponent
\begin{equation}\label{eq:theta}
\theta(\gamma)=\frac{d\ln\mu}{d\ln n}=\begin{cases}0,&\gamma<a-2,\\ \frac{\gamma-a+2}{a-1},&a-2<\gamma<a-1,\\ \frac{1}{a-1},&\gamma>a-1,\end{cases}
\end{equation}
and at $\gamma_2=a-1$ the forcing condenses on a few nodes. The participation ratio
$Y_2=\sum_i\pi_i^2$ tends to zero below $\gamma_2$ and to $1-(a-1)/\gamma$ above it, which is the
Poisson--Dirichlet value \cite{derrida1997random,pitman1997two,filiasi2014condensation}.

\begin{figure}[t]
\includegraphics[width=\columnwidth]{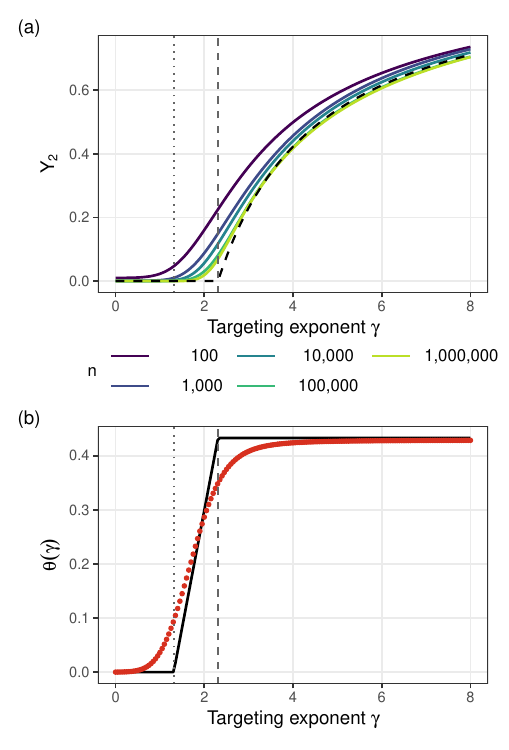}
\caption{Targeting transition for $n=10^2$ to $10^6$ entry nodes with degrees drawn
from the survey law with an uncapped tail ($a=3.31$). (a) Participation ratio $Y_2$
against $\gamma$, with the $n\to\infty$ limit $1-(a-1)/\gamma$ (black dashed). (b) Growth exponent
$\theta(\gamma)$ of the targeted degree fitted over $n=10^3$ to $10^6$ (points) and
Eq.~\eqref{eq:theta} (line). Vertical lines mark $\gamma_1=a-2$ (dotted) and $\gamma_2=a-1$
(dashed).}
\label{fig:transition}
\end{figure}

The two critical points have an elementary origin. The targeted degree
$\mu=\overline{k^{\gamma+1}}/\overline{k^{\gamma}}$ is a ratio of two empirical means over the $n$ entry
nodes. For nonnegative i.i.d.\ variables, the empirical mean converges to a
finite limit if the expectation is finite (strong law of large numbers) and
grows without bound if it is infinite (its converse) \cite{durrett2019probability}.
With tail $p(k)\sim k^{-a}$, $\E[k^{\gamma+1}]$ is finite only for $\gamma<a-2=\gamma_1$ and $\E[k^\gamma]$ only
for $\gamma<a-1=\gamma_2$. Above $\gamma_1$ the numerator therefore grows with $n$ at the rate of
its largest term, $k_{\max}^{\gamma+1}/n\sim n^{(\gamma+1)/(a-1)-1}$, and above $\gamma_2$ the partition
function does the same, which gives the three branches of Eq.~\eqref{eq:theta}.
The same mechanism produces the critical points of sums of random exponentials
\cite{benarous2005limit}. Figure~\ref{fig:transition} confirms both predictions
for the survey law ($a=3.31$) with $n$ up to $10^6$. The fitted plateau of $\theta$ is
0.428 against $1/(a-1)=0.433$, and $Y_2$ approaches its limit slowly and without
self-averaging, as expected near and above a condensation point. Pure power
laws with $a=2.8$, 3.31 and 4 give the same agreement, with plateaus within 3\%
of $1/(a-1)$, and curves for different $n$ collapse against $(\gamma-\gamma_2)\ln n$
\cite{SM}. A degree cap removes the transition (Sec.~\ref{sec:cross}).

\subsection{Depletion and the optimal targeting exponent}
\begin{figure*}[t]
\includegraphics[width=\textwidth]{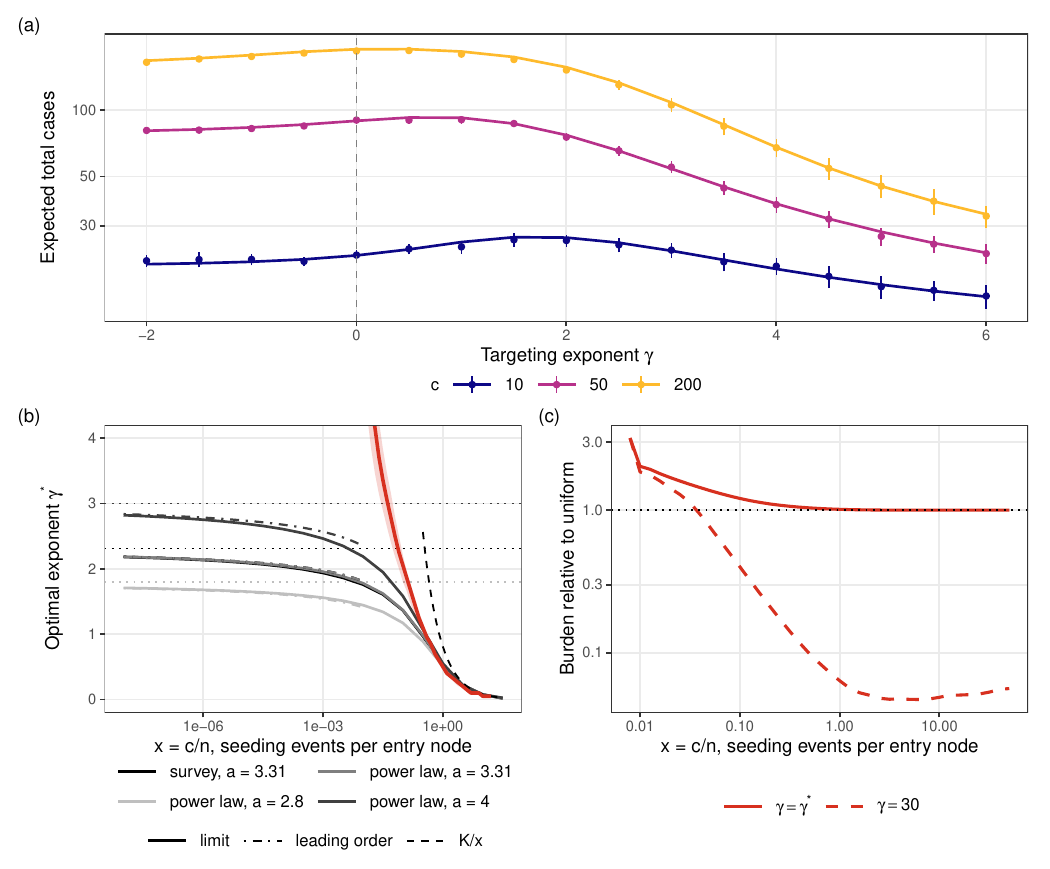}
\caption{Repeated forcing with depletion. (a) Expected total cases against $\gamma$
on survey-calibrated networks at $\tau=0.6$ for $c=10$, 50 and 200. Points show
simulations over 50 networks ($\pm2$ standard errors) and lines the theory. (b)
Optimal exponent $\gamma^*$ against $x=c/n$. The red curve and band are the median and
interquartile range over 100 survey-calibrated networks, grey and black curves
the $n\to\infty$ limits for uncapped laws (solid; $a=2.8$ has infinite $\kappa_1$ and
applies only to the general importance model of Sec.~\ref{sec:gen}), the dot-dashed curves the
leading order $\gamma_2-(a-1)/\ln(1/x)$ of Eq.~\eqref{eq:lead} for $x\le10^{-2}$, the
dashed curve the large-$x$ asymptote $K/x$ and dotted lines $\gamma_2$. (c) Burden at $\gamma^*$ and at $\gamma=30$ relative to
uniform forcing.}
\label{fig:gstar}
\end{figure*}

Under repeated forcing, entry node $i$ is ever seeded with probability
$1-e^{-c\pi_i}$, so the expected numbers of realized seedings and of cases are
\begin{align}
I(\gamma)&=\sum_{i\in B}\left(1-e^{-c\pi_i}\right),\label{eq:IC}\\
C(\gamma)&=\sum_{i\in B}\left(1-e^{-c\pi_i}\right)(1+Ak_i).\label{eq:C}
\end{align}
Uniform forcing maximizes $I$ (Jensen's inequality), but
$dC/d\gamma|_{\gamma=0}=c\,e^{-c/n}A\,\mathrm{Cov}_B(\ln k,k)>0$ and $C$ does not decrease for $\gamma\le0$
\cite{SM}, so the burden-maximizing exponent $\gamma^*$ is strictly positive for every
forcing level. Concentrating the forcing on few nodes wastes it on nodes
already seeded (Fig.~\ref{fig:gstar}a), which bounds $\gamma^*$ in large populations
(Sec.~\ref{sec:asym}). In small populations under weak forcing the burden can
instead keep increasing up to complete concentration. In simulations of
survey-calibrated networks $\gamma^*\approx1.5$--2, 0.5--1 and 0.5 for $c=10$, 50 and 200, and
targeting with $\gamma=30$ lowers the burden by a factor of 10 to 20 at large $c$.

\subsection{Asymptotics of the optimal exponent}\label{sec:asym}
We take $n\to\infty$ at fixed $x$, which describes forcing that grows with the pool
of entry nodes. At fixed total forcing $c$ instead, $x\to0$ and the optimum
approaches $\gamma_2$ (see below). The limit keeps the attack fraction finite, so it
carries the overlap error discussed in Sec.~\ref{sec:model}. In this limit $C/n\to\E[(1-e^{-xk^\gamma/\E k^\gamma})(1+Ak)]$ for
$\gamma<\gamma_2$, and $C/n\to0$ for $\gamma>\gamma_2$, so $0<\gamma^*_\infty(x)\le\gamma_2$. For intense forcing,
$\gamma^*\simeq K/x$, where $K$ solves $\E[(\ln k-\avg{\ln k})(1+Ak)k^{-K}]=0$. For $A=2$, $K$ lies between 0.78 and 0.82 for all tails considered. $K$
depends on $A$ and vanishes as $A\to0$, where reaching the largest number of
entry nodes ($\gamma=0$) is optimal. For sparse forcing the burden is
dominated by nodes near the saturation scale $k_s=(\E k^\gamma/x)^{1/\gamma}$, which gives
\begin{equation}\label{eq:G}
\gamma^*_\infty\simeq\arg\max_\gamma\left\{-\nu\left[\ln\tfrac1x+\ln\E k^\gamma\right]+\ln\Gamma(1-\nu)\right\},
\end{equation}
where $\nu=(a-2)/\gamma$. The maximizer does not depend on $A$, and its relevant
maximum is the interior one near $\gamma_2$, because
the tail approximation fails as $\nu\to1$. To leading order
\begin{equation}\label{eq:lead}
\gamma_2-\gamma^*_\infty\simeq\frac{a-1}{\ln(1/x)} .
\end{equation}
Equation~\eqref{eq:lead} matches the exact limit. For $a=3.31$, $\gamma^*_\infty=2.184$
against $2.185$ at $x=10^{-8}$ and $2.048$ against $2.059$ at $x=10^{-4}$, and for $\eta\ge1$
(Sec.~\ref{sec:gen}) the difference is below 0.07 for all $x\le10^{-4}$ and all tails
considered \cite{SM}. Maximizing Eq.~\eqref{eq:G} itself, where its
interior maximum exists (below an $(a,\eta)$-dependent value of $x$), reproduces the
exact limit with errors of $10^{-3}$--$10^{-6}$, so it also captures the subleading
corrections. The approach to the condensation point is therefore
logarithmic in $x$. Finite pools approach this limit from above and only
logarithmically in $n$. For the survey law at $x=10^{-4}$, $\gamma^*=5.95$ for $n=10^4$ and
2.45 for $n=10^6$, against $\gamma^*_\infty=2.03$ and $\gamma_2=2.31$ \cite{SM}. Under sparse forcing
the optimum therefore sits at the condensation point, below it only in very
large pools. For intense forcing the limit holds already at $n=100$ ($\gamma^*=0.55$
against 0.52 at $x=1$).
Values of $K$ and the comparison with the exact limit for all tails and
response exponents are given in the SM \cite{SM}.

\subsection{General importance and response}\label{sec:gen}
\begin{figure}[t]
\includegraphics[width=\columnwidth]{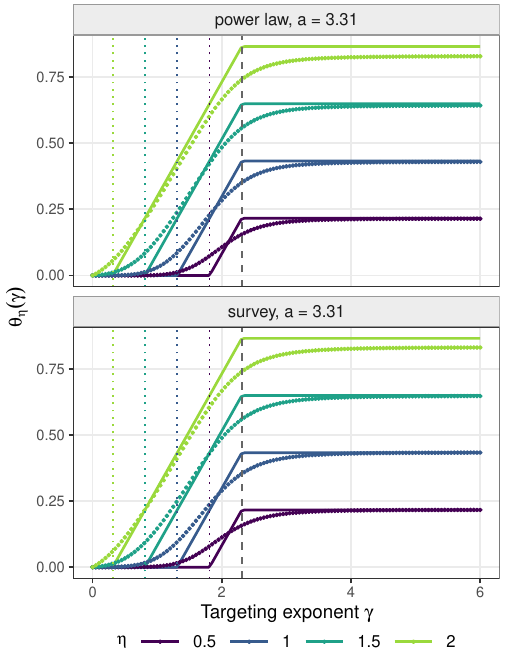}
\caption{General response. Growth exponent $\theta_\eta(\gamma)$ of the targeted response
for $\eta=0.5$, 1, 1.5 and 2, for a power law with $a=3.31$ (top) and the survey law
with an uncapped tail (bottom). Points are fits over $n=10^3$ to $10^6$ and lines
the prediction. The first kink moves with $\gamma_1=a-1-\eta$ (dotted, colored by $\eta$)
and condensation stays at $\gamma_2=a-1$ (dashed).}
\label{fig:gen}
\end{figure}

Nothing above depends on $w$ being a degree. Let entry node $i$ carry any
importance $w_i$ with tail $P(W\ge w)\simeq C_0w^{-(a-1)}$, targeted as $\pi_i\propto w_i^\gamma$, and
let a seeding event at $i$ produce $Aw_i^\eta$ secondary events, $0<\eta<a-1$ (the
degree model is $w=k$, $\eta=1$). The exponent $\eta$ plays the role of the dynamical
exponents that relate the response of a node to its degree in network dynamics
\cite{barzel2013universality}. The targeted response
$\mu_\eta=\overline{w^{\gamma+\eta}}/\overline{w^\gamma}$ diverges at
\begin{equation}\label{eq:g1eta}
\gamma_1=a-1-\eta ,
\end{equation}
while condensation of the forcing remains at $\gamma_2=a-1$ for any response. The
growth exponent is $\theta_\eta=0$ below $\gamma_1$, $(\gamma-\gamma_1)/(a-1)$ between the
transitions and $\eta/(a-1)$ above $\gamma_2$. Simulations with $\eta=0.5$, 1, 1.5 and 2
reproduce the shift of the first kink with $\eta$ and the plateaus $\eta/(a-1)$ to
within 4\% (Fig.~\ref{fig:gen}). The positivity of $\gamma^*$, the large-$x$ asymptote
(with $1+Ak$ replaced by $1+Aw^\eta$; $K\approx0.37$, 0.8 and 1.7 for $\eta=0.5$, 1 and 2)
and Eq.~\eqref{eq:lead} carry over unchanged, so the phase structure is set by
the tail of the importance distribution and the response exponent only.

\subsection{Cut-off crossover and real populations}\label{sec:cross}
\begin{figure}[t]
\includegraphics[width=\columnwidth]{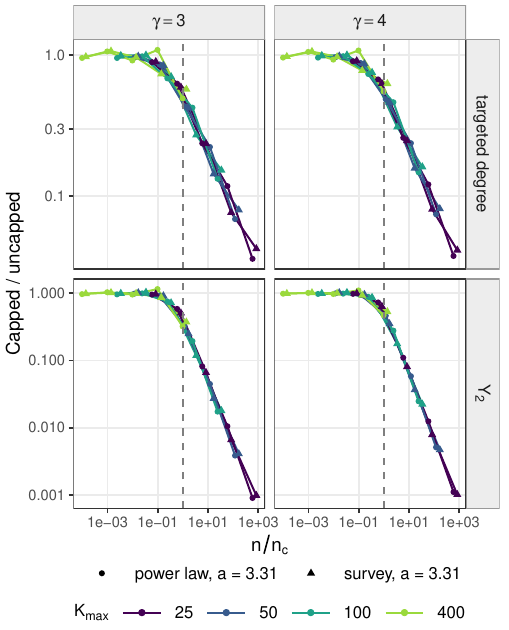}
\caption{Cut-off crossover. Ratios of capped to uncapped targeted degree (top)
and participation ratio (bottom) against $n/n_c$, with
$n_c=(K_{\max}/k_0)^{a-1}/P_{\mathrm{tail}}$, for caps $K_{\max}=25$ to 400 (colors) and two tail
laws (symbols), at $\gamma=3$ and 4, both above $\gamma_2$. The dashed line marks $n=n_c$.}
\label{fig:cross}
\end{figure}

A finite cut-off $K_{\max}$ removes the transition. The cut-off is irrelevant while
the largest expected degree among $n$ entry nodes is below $K_{\max}$, that is, for
$n\ll n_c=(K_{\max}/k_0)^{a-1}/P_{\mathrm{tail}}$, where the tail is $P(K\ge k)=P_{\mathrm{tail}}(k/k_0)^{-(a-1)}$
for $k\ge k_0$. For the sexual-partnership distribution used below
($k_0=4$, $P_{\mathrm{tail}}=0.0575$, $a=3.31$, $K_{\max}=25$), $n_c\approx1.2\times10^3$, so entry-node pools
of $10^2$--$10^3$ sit in the rounded, finite-size regime of the transition.
For caps $K_{\max}=25$--400 and two tail laws, the ratios of capped to uncapped
targeted degree and participation ratio are close to 1 for $n/n_c\lesssim0.05$, fall
to 0.25--0.5 near $n/n_c=1$ and vanish for $n\gg n_c$, collapsing as functions of
$n/n_c$ (Fig.~\ref{fig:cross}). The entry-node pools of the networks used below
(about 90 and 940 nodes) lie at $n/n_c\approx0.08$, where the cut-off barely
affects the finite-size behavior, and at $n/n_c\approx0.8$, in the middle of the
crossover. The position of a
real population on this crossover therefore determines how strongly the
targeting transition affects it.

\subsection{Importation into sexual contact networks}
To locate a real system on this phase diagram we built bipartite networks whose
degrees are numbers of opposite-sex partners in the past year from a national
survey \cite{chandra2011sexual}, with a power-law tail for four or more
partners \cite{liljeros2001web}, and entry nodes given by bisexual men with at
least one female partner \cite{copen2016sexual} (details in \cite{SM}). Across
100 realizations $\sqrt{\kappa_1\kappa_2}\le0.92$, so these networks are subcritical at
every transmissibility. Each seeding event produces a finite cluster, and the
results of Secs.~\ref{sec:results}A--D apply directly. With $5\times10^3$ people of
each sex there are about 90 entry nodes, most with a single partner. The
targeting gain saturates at $k_{\max}/\mu(0)\approx5$ (Fig.~\ref{fig:gain}), $\gamma^*$ follows the
depletion law of Fig.~\ref{fig:gstar}, and the pool lies at $n/n_c\approx0.08$
(Fig.~\ref{fig:cross}). A pool this small is far from the asymptotic regime, so
the transitions appear only as the finite-size behavior of Figs.~\ref{fig:gain}
and \ref{fig:gstar}.
Whether the tail of partner numbers is a power law is debated
\cite{jones2003assessment,handcock2004likelihood}. The results depend on it
only through $a$, and for pools this small the cut-off, not the tail law,
limits the response.

Household surveys miss people with very many partners. Closing the network
with a small core of sex workers calibrated to independent data
\cite{brewer2000prostitution} makes large outbreaks possible and adds a
feature that degree alone does not capture. About 5\% of entry nodes are
clients of the core. A single seeding event into one of them
reaches 1\% of the population with probability 0.50 at $\tau=0.3$, against 0.003
for other entry nodes, and under repeated forcing ($\tau=0.3$, $c=10$) the
probability of such an outbreak rises from 0.03 at $\gamma=-2$ to 0.26 at $\gamma=0$ and
0.91 at $\gamma=2$ \cite{SM}. In this regime targeting decides whether a large
outbreak occurs, not only its size. The core is small and loop-rich, so the
tree-like theory does not describe it. Its systematic analysis is left to
future work.

\section{Discussion}\label{sec:discussion}
When external forcing enters a heterogeneous network repeatedly, the burden
depends on how the forcing is distributed over entry nodes and not only on its
amount. The mechanism has three parts. Each seeding event produces a response that grows
with the importance of the seeded node, which favors concentration. The
targeting measure is a Gibbs measure whose condensation point $\gamma_2=a-1$ is set by
the tail of the importance distribution. And depletion penalizes
concentration, because forcing aimed at nodes already reached is wasted. In large
populations the balance gives a finite optimal exponent bounded by $\gamma_2$, decaying as $K/x$ for
intense forcing and approaching $\gamma_2$ logarithmically for sparse forcing. Finite
pools under sparse forcing sit at or above $\gamma_2$.
Uniform forcing reaches the largest number of distinct entry nodes but never
maximizes the cascade.
Depletion is the stochastic counterpart of the redundancy between seeds that
makes influence maximization submodular \cite{kempe2003maximizing}. Here it
can be computed in closed form and set against the gain from concentration.

Nothing in this chain is specific to epidemics. Any process in which seeding
events arrive repeatedly at a finite set of entry points, each event produces
a response growing as a power of the entry point's importance, and an entry
point can be reached only once, has the same structure. Examples are the
seeding of information or products in social networks, the introduction of
invasive species through ports, and failures entering infrastructure networks.
The tail exponent of the importance distribution and the response exponent
determine the critical points, and the number of seeding events per entry
point determines where between uniform targeting and condensation the most
damaging allocation lies.

The bipartite structure enters the results only through the constant $A$ of
Eq.~\eqref{eq:sk}. In a unipartite configuration network with mean excess
degree $\kappa$, setting $\kappa_1=\kappa_2=\kappa$ gives $A=\tau/(1-\tau\kappa)$ and $R=\tau\kappa$, so the
response is again linear in the degree of the seeded node. The relative gain
$\mu(\gamma)/\mu(0)$, the transitions at $\gamma_1$ and $\gamma_2$, the positivity of $\gamma^*$ and
Eq.~\eqref{eq:lead} do not depend on the value of $A$ and carry over unchanged,
and the large-$x$ constant $K$ changes only through the new value of $A$. Two
differences arise when every node can be an entry node, as in a scale-free
network without a distinguished bridge class. For $2<a<3$, $\kappa$ diverges with
network size and the epidemic threshold vanishes
\cite{pastorsatorras2015epidemic}, so the linear response holds only below a
degree cut-off, while the targeting measure and the depletion law, which do
not involve $\kappa$, hold for any tail. In addition, clusters started by
different seeding events can reach other entry nodes, which adds internal
depletion to the external depletion of the model. The no-overlap approximation
of Sec.~\ref{sec:model} then limits the results to small attack fractions.

Real populations are finite and their heterogeneity is bounded. The crossover
size $n_c$ makes this quantitative. Populations with $n\ll n_c$ show the
transition rounded by finite size, while for $n\gg n_c$ the cut-off removes it.
The survey-calibrated networks we studied lie below or near $n_c$, so their
response shows the finite-size form of the transitions rather than the
asymptotic one.

The analysis has limitations. The tree-like approximation fails close to the
epidemic threshold and for small, densely shared cores, and clusters started
by different seeding events are assumed not to overlap, which overestimates
the burden by about 10\% when forcing is intense and transmissibility high.
The network is static, and whether a node is an entry node is independent of
its degree. Correlations between entry-node status and degree, temporal
partnership dynamics and loop-rich cores are left for future work. Finally, we
optimize the expected burden. Cluster sizes are broad near threshold, so the
median burden or the probability of a large outbreak can have a different
optimum. Above threshold, the probability of at least one major outbreak also
peaks at an intermediate $\gamma$ \cite{SM}.

\begin{acknowledgments}
The author thanks Tarleton State University for its support. The author used
Claude Opus 5.5 (Anthropic) to assist with simulation code, asymptotic
calculations, the schematic in Fig.~\ref{fig:schematic}, and text editing. All
derivations were checked by the author and against simulations, and all code
outputs were validated against the analytical predictions. The author takes
full responsibility for the content.
\end{acknowledgments}

\section*{Data Availability}
The code, simulation outputs and figure scripts that support the findings of
this article are openly available at
\href{https://github.com/jdiestra1977/degree-targeted-importation}{\mbox{\nolinkurl{https://github.com/jdiestra1977/}}}\hfil\penalty0\hfilneg
\href{https://github.com/jdiestra1977/degree-targeted-importation}{\mbox{\nolinkurl{degree-targeted-importation}}}. The degree distributions are calibrated to published
survey data \cite{chandra2011sexual,copen2016sexual,liljeros2001web,brewer2000prostitution},
and the forcing profile uses public incidence data \cite{nycdohmh2022mpox}.

\bibliography{main}

\end{document}